\documentclass[twocolumn,prl,showpacs]{revtex4}%
\usepackage{graphicx}%
\usepackage{amsmath}%
\usepackage{amsfonts}%
\usepackage{amssymb}
\usepackage{bm}

\def\k{{ {\bm k} }}

\def\q{{ {\bm q} }}

\def\w{{\omega}}
\def\a{{\alpha}}

\allowdisplaybreaks[4]

\begin{document}
\title{
 In-plane anisotropy of transport coefficients in the electronic nematic
 states: Universal origin of the nematicity in Fe-based superconductors
}
\author{Seiichiro \textsc{Onari}$^{1,2}$
and Hiroshi \textsc{Kontani}$^{3}$}
\date{\today }

\begin{abstract}
The origin of the electronic nematicity and its remarkable
 material-dependence are famous longstanding unsolved issues in Fe-based superconductors. 
To attack these issues, we focus on the in-plane anisotropy of the resistivity: 
In the nematic state in FeSe, the relation $\rho_x>\rho_y$ holds, 
where $\rho_{x(y)}$ is the resistivity along the longer (shorter) Fe-Fe axis. 
In contrast, the opposite anisotropy $\rho_x<\rho_y$ is realized in
 other undoped Fe-based superconductors. 
Such nontrivial material dependence is naturally explained
 in terms of the strongly orbital-dependent inelastic quasiparticle scattering
realized in the orbital-ordered state.
The opposite anisotropy between FeSe ($\rho_x>\rho_y$) and 
other undoped compounds ($\rho_x<\rho_y$) reflects the 
difference in the number of hole-pockets.
We also explain the large in-plane anisotropy of the thermoelectric power in the nematic state.
\end{abstract}

\address{
$^1$ Department of Physics, Okayama University, Okayama 700-8530, Japan
\\
$^2$ Research Institute for Interdisciplinary Science, Okayama University, Okayama 700-8530, Japan
\\
$^3$ Department of Physics, Nagoya University, 
Furo-cho, Nagoya 464-8602, Japan
}
 
\pacs{74.70.Xa, 75.25.Dk, 72.10.-d, 72.15.Jf} 

\sloppy

\maketitle

The emergence of the electronic nematic states below the structure
transition temperature $T_{\rm S}$ is one of the significant universal
features in Fe-based superconductors. However, the realized electronic
properties exhibit remarkable compound dependences. One example is the absence
of magnetism in FeSe and the presence of magnetism in the nematic states
(N\'eel temperature $T_{\rm N} \lesssim T_{\rm S}$) in other compounds. 
As possible nematic order parameters, the
spin-nematic order \cite{Fernandes,Fernandes2} and the orbital order \cite{Onari-SCVC,Onari-Hdoped,Kruger,PP,WKu} have been studied intensively so far.
Recently, the present authors explained the nematicity without
magnetization in FeSe as the orbital order caused by the Aslamazov-Larkin vertex correction \cite{FeSe-Yamakawa2}. 
The current fundamental question is whether the origin of the nematicity
is universal or material-dependent
\cite{FeSe-Yamakawa2,Chubukov-PRX}.


To answer this question,
the strong in-plane anisotropy of transport coefficients has been studied 
intensively as a key electronic property in the nematic state
\cite{Ba122-aniso-rho,Ba122-aniso-rho2,Jiang,FeSe-aniso-rho,FeSe-aniso-rho2,Impurity-rho-aniso-theory,Inoue,Nematogen-exp,Nematogen-theory,Tohyama,Fernandes2}.
In  Ba(Fe$_{1-x}$Co$_x$)$_2$As$_2$, Ba(As$_{1-x}$P$_x$)$_2$ and EuFe$_2$(As$_{1-x}$P$_x$)$_2$, 
large $C_2$ anisotropy in the resistivity $\Delta\rho\equiv\rho_x-\rho_y <0$
appears in detwinned samples below $T_{\rm S}$, where $\rho_\mu$ is the resistivity along the
$\mu$-axis \cite{Ba122-aniso-rho,Ba122-aniso-rho2,Jiang}. 
The relation $\Delta\rho<0$ is observed in the non-magnetic nematic state for
$T_{\rm S} > T > T_{\rm N}$, and even for $T \gtrsim T_{\rm S}$ 
under the weak uniaxial stress. 
Remarkably, the opposite anisotropy $\Delta\rho>0$
is realized in FeSe \cite{FeSe-aniso-rho,FeSe-aniso-rho2}. 
According to these observations, one may expect that 
the origin of nematicity in FeSe is special.

The anisotropic elastic scattering due to the impurity-induced $C_2$
local orbital order (orbital
nematogen) \cite{Impurity-rho-aniso-theory,Inoue} and the
magnetic nematogen \cite{Nematogen-exp,Nematogen-theory}, 
and the anisotropic quasiparticle velocity \cite{Tohyama}
have been discussed. 
On the other hand,
the anisotropic inelastic scattering due to the $C_2$ spin fluctuations
was discussed based on the spin-nematic scenario \cite{Fernandes2}. 
In BaFe$_2$As$_2$ \cite{Ba122-aniso-rho2},
the anisotropy of resistivity is reduced after the annealing,
indicating that both elastic scattering and inelastic one contribute to
the anisotropy in BaFe$_2$As$_2$.
In contrast to Ba122 compounds, $\rho_\mu$ in FeSe exhibits sizable 
anisotropy even in the clean limit samples, in which
the elastic scattering is negligible at $T\sim T_{\rm S}$ (=90K).
Therefore, the in-plane resistivity anisotropy in FeSe below $T_{\rm S}$
should originate from the inelastic scattering. 
The opposite anisotropic relation between FeSe ($\Delta\rho>0$) and
other compounds ($\Delta\rho<0$)
provides us a crucial hint to understand the origin of
the nematicity in Fe-based superconductors.

In this paper, we study the
in-plane anisotropy of resistivity and thermoelectric power (TEP)
below $T_{\rm S}$ based on the orbital-order scenario.
Under the nematic orbital order, the spin susceptibility 
becomes strongly orbital-dependent, so the total spin susceptibility 
possesses large $C_2$ anisotropy: $\chi^s(\pi,0)\gg\chi^s(0,\pi)$ \cite{Kontani-quad}.
Then, the inelastic scattering rate on band $b$, $\gamma^b_{\k}$,
possesses strong in-plane anisotropy 
due to the orbital-dependent spin fluctuations.
By taking this fact into account,
the characteristic anisotropy of the transport coefficients
in the nematic states are naturally understood.
In particular, the anisotropy $\Delta\rho>0$ characteristic in FeSe
originates from the ``singleness of the hole pocket''.
This study leads to the conclusion that the
orbital nematicity is universal in various Fe-based superconductors.

The nematic orbital order below $T_S$
is given by the vertex correction (VC), which represents the 
many-body effects beyond the random-phase-approximation (RPA)
 \cite{Onari-SCVC,Onari-Hdoped,FeSe-Yamakawa2,Onari-Springer}.
Based on this self-consistent vertex correction (SC-VC) theory,
we can explain the strong orbital fluctuations, which are measured 
by the softening of $C_{66}$ and Raman study \cite{Kontani-Yamakawa-Raman},
and the ``sign-reversing orbital polarization in $\k$-space'' below
$T_{\rm N}$ in FeSe \cite{Onari-FeSe}.
This attractive orbital-order scenario is confirmed by the present study
for various Fe-based superconductors.

We set the $x$ and $y$ axes parallel to the nearest Fe-Fe
bonds, and denote the orbital $d_{3z^2-r^2}$, $d_{xz}$, $d_{yz}$, $d_{xy}$, and
$d_{x^2-y^2}$ as $l=1$, 2, 3, 4, and 5, respectively.
We employ the eight-orbital {\it d-p} Hubbard model \cite{FeSe-Yamakawa2,Onari-FeSe}
based on the first-principles calculation
\begin{eqnarray}
H_{\rm M}(r)=H_{\rm M}^0+rH_{\rm M}^U+H_{\rm M}^{\rm orb} \ \ \
\mbox{(M = LaFeAsO, FeSe)},
\label{eqn:Ham}
\end{eqnarray}
where 
$H_{\rm M}^0$ is the eight-orbital tight-binding model, and $H_{\rm M}^U$ is the first-principles screened Coulomb potential
for $d$-orbitals in Ref.  \cite{Arita2}. 
The factor $r(<1)$ is the parameter introduced to adjust the spin fluctuation
strength.
$H_{\rm M}^{\rm orb}=\sum_{\bm{k},l=2,3}\Delta E_l(\bm{k})n_l(\bm{k})$
is given by the $\bm{k}$-dependent orbital-polarization energy $\Delta E_l(\bm{k})$ and
the electron density for $l$ orbital $n_l(\bm{k})$. $\Delta E_l(\bm{k})$
becomes $0$ for $T\ge T_{\rm S}$. In the LaFeAsO model, we employ the
constant orbital polarization $\Delta E_{xz}(\bm{k})=-\Delta_E$ and $\Delta E_{yz}(\bm{k})=\Delta_E$. In the FeSe model, we employ the sign reversing
orbital polarization $\Delta E_{xz(yz)}(\bm{k})$ obtained in the
previous microscopic
study \cite{Onari-FeSe}, which is consistent with angle-resolved
photoemission spectroscopy (ARPES) measurements \cite{FeSe-ARPES62}. 
Here, the relation $\Delta E_{xz}(k_x,k_y)=-\Delta E_{yz}(k_y,k_x)$
holds, and the maximum
orbital polarization is given by $\Delta_E=\Delta E_{yz}$(X)$=-\Delta E_{xz}$(Y).
See the the Supplemental Material (SM), Sec. A \cite{SM} for details.

In the presence of $\Delta E_l(\k)$, we calculate the
 spin (orbital) susceptibilities $\hat{\chi}^{s(c)}(q)=\hat{\chi}^{\rm
 irr}(q)/[1-\hat{\Gamma}^{s(c)}\hat{\chi}^{\rm irr}(q)]^{-1}$ using the RPA, where $\chi_{ll',mm'}^{\rm
 irr}(q)=-\frac{T}{N}\sum_k{G}_{l,m}^0(k+q){G}_{m',l'}^0(k)$ is the
 irreducible susceptibility in the orbital basis, and $\hat{\Gamma}^{s(c)}$
is the bare Coulomb interaction \cite{Kontani-quad0}. $\hat{G}^0$ is the Green function matrix without the self-energy.
We denote $k=(\bm{k},\epsilon_n)$ with fermion Matsubara frequency
$\epsilon_n=(2n+1)\pi T$, and $q=(\bm{q},\omega_n)$ with boson Matsubara
frequency $\omega_n=2n\pi T$. The spin Stoner factor $\a_s$ is defined as
 the maximum eigenvalue of $\hat{\Gamma}^s\hat{\chi}^{\rm irr}(\q)$.  At $T=T_{\rm N}$,
 $\a_s=1$ is satisfied.
We also calculate the self-energy matrix $\hat{\Sigma}(k)=\frac{T}{N}\sum_q\hat{V}^\Sigma(q)\hat{G}(k-q)$, where
$\hat{G}$ is the Green function matrix,
and $\hat{V}^\Sigma$ is the interaction matrix
for the self-energy \cite{Onari-Hdoped,Onari-FeSe,Onari-Springer}.
We employ the RPA for $\hat{V}^\Sigma$, and calculate $\hat{G}=[(\hat{G}^0)^{-1}-\hat{\Sigma}]^{-1}$ and
$\hat{\Sigma}$ self-consistently.
Details of the formulation are described in the SM, Sec. A \cite{SM}.
Qualitatively similar results are obtained from
the  fully self-consistent approximation by including
the self-energy in $\hat{V}^\Sigma$.
Hereafter, we take $N=N_x\times N_y=128\times128$ $\bm{k}$ meshes, 
 1024 Matsubara frequencies, and $T=20$meV unless otherwise noted.

\begin{figure}[!htb]
\includegraphics[width=\linewidth]{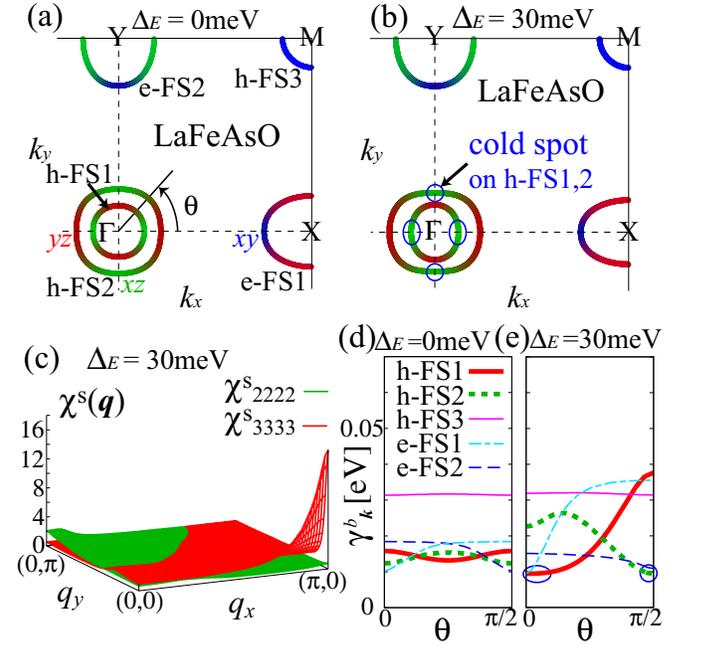}
\caption{
(a) The hole-like FSs (h-FS1-3) and the electron-like FSs (e-FS1,2) in the LaFeAsO model for $\Delta_E=0$ and
 (b) those for $\Delta_E=30$meV, where $\theta$ denotes the azimuthal angle
 on a FS ($\theta=0$ corresponds to the $k_x$ direction). 
The colors correspond to 2 (green), 3 (red), and 4 (blue), respectively.
(c) $\bm{q}$ dependencies of $\chi^s_{22,22}(\q)$ and $\chi^s_{33,33}(\q)$ for
 $\Delta_E=30$meV ($\a_s=0.967$).
(d) $\theta$ dependences of $\gamma^b_{\k}$ on the FSs for $\Delta_E=0$
 and (e) those for $\Delta_E=30$meV. Cold spots on the h-FS1,2 are marked by
 blue circles in (b) and (e). 
}
\label{fig1}
\end{figure}

We start with the LaFeAsO model. Its bandstructure is similar to that of
Eu122 and Ba122.
Figure \ref{fig1}(a) shows the Fermi surfaces (FSs) for $\Delta_E=0$, where the hole-FSs are denoted as h-FS1-3, and
the electron-FSs are denoted as e-FS1,2.
Figure \ref{fig1}(b) shows the deformed
FSs for $\Delta_E=30$meV.
Here, the orbital splitting $2\Delta_E$ is comparable to 
 the ARPES measurement in
BaFe$_2$As$_2$ \cite{ARPES-Shen,Shimojima} for $T\ll T_{\rm
N}$.
We put $r=0.334$, in which the spin Stoner factor
$\alpha_s$ is $0.898$ for $\Delta_E=0$. Then, $\alpha_s$ increases to $0.990$ when
$\Delta_E=50$meV. 
Figure \ref{fig1}(c) shows the spin susceptibility for
$\Delta_E=30$meV, in which the relation $\chi^s_{33,33}(\pi,0)\gg
\chi^s_{22,22}(0,\pi)$ gives the prominent $C_2$ anisotropic spin
susceptibility $\chi^s(q)\equiv\sum_{l,m}\chi^s_{ll,mm}(q)$.
Such strong orbital dependence in $\chi^s$ causes the
orbital-dependent quasiparticle damping $\gamma^b_{\k}(=-{\rm
Im}\Sigma^b(\bm{k},+i0))$ as shown in Fig. \ref{fig1}(e). The cold spot is defined as the position on the FS with
minimum value of $\gamma^b_{\bm{k}}$. Since
the spin fluctuations mainly develop in the $d_{yz}$ orbital for
$\Delta_E>0$, 
the cold spots are located on the FS composed of the $d_{xz}$ orbital. In Fig. \ref{fig1}(b), we show only the cold spots on
the h-FS1,2 since they are significant for the $C_2$ transport phenomena.
The anisotropy in the transport coefficients is determined by the
positions of the cold spots.
Note that, in the present study, we ignore the damping due to the orbital fluctuations caused by the VC. However, the positions of cold 
spots are unchanged by the orbital fluctuations since only
 $\chi^c_{33,33}$ is enhanced by the VC \cite{Yamakawa3}.
Therefore, the anisotropy in the transport coefficients obtained in this
study is expected to be unchanged. This is our important future issue.

\begin{figure}[!htb]
\includegraphics[width=\linewidth]{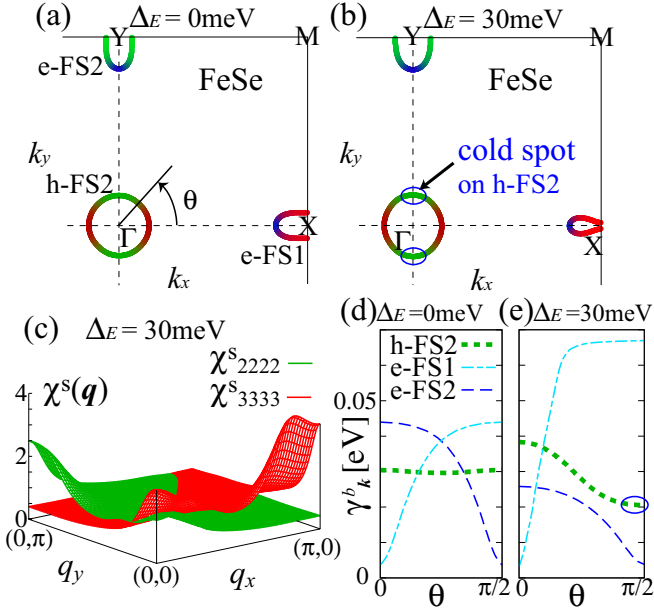}
\caption{
(a) FSs of the FeSe model for $\Delta_E=0$ and (b) those for $\Delta_E=30$meV. 
Here, the h-FS1 and h-FS3 are absent. (c) $\bm{q}$ dependencies of
 $\chi^s_{22,22}(\q)$ and $\chi^s_{33,33}(\q)$ for $\Delta_E=30$meV ($\a_s=0.867$).
(d) $\theta$ dependences of $\gamma^b_{\k}$ on the FSs for $\Delta_E=0$
 and (e) those for $\Delta_E=30$meV. The cold spots on the h-FS2 are marked
 by blue circles.
}
\label{fig2}
\end{figure}

Next, we move to the FeSe model. We introduce the mass
enhancement factor $z^{-1}_{xy}=1.6$ by following
Refs.  \cite{FeSe-Yamakawa2,Onari-FeSe}.
In Figs. \ref{fig2}(a) and \ref{fig2}(b), the FSs for $\Delta_E=0$meV
and the FSs for $\Delta_E=30$meV are shown, respectively.
The h-FS1 and h-FS3 are absent in the present FeSe model \cite{FeSe-model}.
We put $r=0.218$, where $\alpha_s$ is $0.846$ for $\Delta_E=0$meV. Then,
$\alpha_s$ increases to $0.870$ when $\Delta_E=50$meV.  As shown in Fig. \ref{fig2}(c), the spin susceptibilities for
$\Delta_E=30$meV have
the orbital-dependent $C_2$ anisotropy $\chi^s_{33,33}(\pi,0)>\chi^s_{22,22}(0,\pi)$.
Figures \ref{fig2}(d) and \ref{fig2}(e) show the momentum
dependences of $\gamma^b_{\k}$ on the FSs for $\Delta_E=0$meV and those
for $\Delta_E=30$meV, respectively. 
In Figs. \ref{fig2}(b) and
\ref{fig2}(e), we show the cold spots on the h-FS2, which play an
important role for the $C_2$ transport phenomena.

Next, we study the resistivity $\rho$ due
to the strongly anisotropic inelastic scattering.
Using the linear response theory, the conductivity $\sigma_{\mu}$ along the $\mu(=x,y)$ direction is obtained by
\begin{eqnarray}
\sigma_{\mu}=\frac{e^2}{N} \sum_{\bm{k},b}\int_{-\infty}^{\infty}
 \frac{d\omega}{\pi}\left(-\frac{\partial f(\omega)}{\partial \omega}\right)
 \left|v^\mu_{b,\k} G^b_{\k}(\w+i0)\right|^2 ,
\end{eqnarray}
where $-e$ is the charge of an electron,
and $f(\omega)$ is the Fermi distribution function.
$v^{\mu}_{b,\k}=\frac{\partial \varepsilon^b_\k}{\partial k_{\mu}}$ is the velocity along the $\mu$ direction, where $\varepsilon^b_\k$ is the
dispersion of band $b$.
 $G^b_{\k}(\w+i0)$ denotes the retarded Green function.
In this study, we neglect the VC for the current, since its effect is
small for $\rho$ and the
TEP \cite{Kontani-S,Kontani-S2,Kontani-Review,Benfatto-RH-CVC}, whereas it is important
for the Hall
coefficient and magnetoresistance \cite{Kontani-S,Kontani-S2,Kontani-Review,Benfatto-RH-CVC}. The
study of the current VC is our important future issue.

\begin{figure}[!htb]
\includegraphics[width=\linewidth]{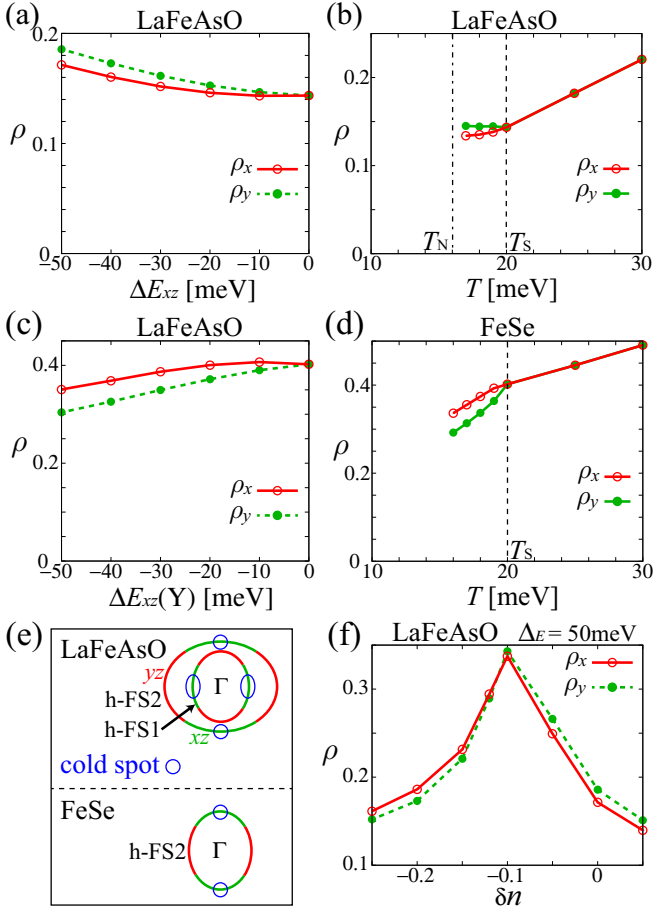}
\caption{
(a) $\Delta E_{xz}$ dependence of $\rho_{\mu}$, and (b) $T$ dependence
 of $\rho_{\mu}$ in the LaFeAsO model.
(c) $\Delta E_{xz}$(Y) dependence of $\rho_{\mu}$, and (d) $T$ dependence of $\rho_{\mu}$ in the FeSe model.
$\rho=1$ corresponds to $(\hbar a_c)/e^2\sim 250\mu$cm$\Omega$
for the interlayer distance $a_c=0.6$nm.
(e) Schematic figures of FSs with the cold spots around the $\Gamma$
 point. (f) Carrier doping $\delta n$ dependence of $\rho_{\mu}$ for
 $\alpha_s=0.990$ and $\Delta_E=50$meV in the LaFeAsO
 model. 
}
\label{fig3}
\end{figure}

Figure \ref{fig3}(a) shows the resistivity $\rho_\mu=1/\sigma_\mu$
obtained for $\Delta E_{xz}=-50$-$0$meV in the LaFeAsO model at $T=20$meV.
 We also show the $T$ dependence
of $\rho_{\mu}$ in the LaFeAsO model in Fig. \ref{fig3}(b) by assuming the $T$ dependence of $\Delta_E$ as the mean-field-like behavior
$\Delta_E=\Delta_E^0\tanh(1.74\sqrt{T_{\rm S}/T-1})$. Here, we put
$\Delta_E^0=50$meV and $T_{\rm S}=20$meV. Then, we obtain $T_{\rm
N}=16$meV from the condition $\a_s=1$.
The obtained in-plane anisotropy
$\Delta\rho<0$ below $T_{\rm S}$ is
consistent with the experimental results in
Ba122 \cite{Ba122-aniso-rho,Ba122-aniso-rho2} and Eu122 \cite{Jiang}.
In contrast, in Figs. \ref{fig3}(c) and \ref{fig3}(d), the opposite in-plane anisotropy $\Delta\rho>0$ is
obtained in the FeSe model.
This result is also consistent with the
experiments in FeSe \cite{FeSe-aniso-rho,FeSe-aniso-rho2}.

Here, we explain why the obtained in-plane anisotropy of
resistivity is opposite between
the FeSe model and the LaFeAsO
model.
In both systems, the anisotropy of $\rho$ mainly stems from the
hole-pockets h-FS1,2, of which the schematic figures are shown in Fig. \ref{fig3}(e). Since the Fermi velocity on the cold spots on the h-FS1 (h-FS2) is
parallel to $k_x$-axis ($k_y$-axis), the h-FS1 (h-FS2) contributes to the
relation $\Delta\rho<0$ $(\Delta\rho>0)$.
In the LaFeAsO model, the relation $\Delta\rho<0$ is realized since the area of
the cold spot on the h-FS1 around $\theta\sim 0$ is very wide as shown in
Figs. \ref{fig1}(b) and \ref{fig1}(e).
In contrast, in the FeSe model, the opposite relation $\Delta\rho>0$ is realized by the
cold spots on the h-FS2 since the h-FS1 is absent.

We verified that the e-FSs are not essential for the
opposite anisotropy of resistivity between FeSe and LaFeAsO.
In both models, the cold
spots on the e-FSs are located on the $d_{xy}$ orbital region, and the area
of the cold spot on the e-FS1 is narrower than that on the e-FS2 in the
orbital-ordered state due to the strong spin fluctuations on the
$d_{yz}$ orbital: See Figs. \ref{fig1}(e) and \ref{fig2}(e). For this reason, the e-FSs contribute to the
relation $\Delta\rho\gtrsim0$ below $T_{\rm S}$.
In FeSe, both the h-FSs and the e-FSs contribute to the
positive $\Delta\rho$. In LaFeAsO, $\Delta\rho$ is negative since the contribution from the e-FSs are
considerably small. 
Therefore, we conclude that the opposite in-plane anisotropy of
resistivity between FeSe and LaFeAsO originates from the presence or
absence of the inner hole-pocket.

In Fig. \ref{fig3}(f), we also show the carrier doping $(\delta n)$ dependences of the in-plane anisotropy
of $\rho$ in the LaFeAsO model for $\Delta_E=50$meV. For each $\delta
n$, $r$ is adjusted to satisfy $\alpha_s=0.990$ for $\Delta_E=50$meV. In
heavily hole-doped case ($\delta n<-0.12$), $\Delta\rho$ is reversed to
positive since the contribution from the h-FS2 becomes
large, consistently with previous theoretical and experimental reports \cite{Fernandes2,Breitkreiz,hole-doped122,hole-doped122-2}. 
Details are described in the SM, Sec. C \cite{SM}.

The anisotropy $\rho_x\ne\rho_y$ due to the $C_2$ spin
fluctuations has been discussed in terms of the spin-nematic scenario \cite{Fernandes2,Breitkreiz}.
In the present paper, we explained that the orbital dependence of the spin fluctuations, which is ignored in the
spin-nematic theory, is essential to understand the characteristic
difference between FeSe and Ba122. In FeSe, the anisotropy of $\rho$ should
originate from the inelastic scattering since the sample is very clean.
In Ba122, in contrast, the anisotropic elastic scattering (nematogen)
 also gives sizable contribution as discussed in Refs.  \cite{Impurity-rho-aniso-theory,Inoue,Nematogen-exp,Nematogen-theory,Tohyama}.

\begin{figure}[!htb]
\includegraphics[width=\linewidth]{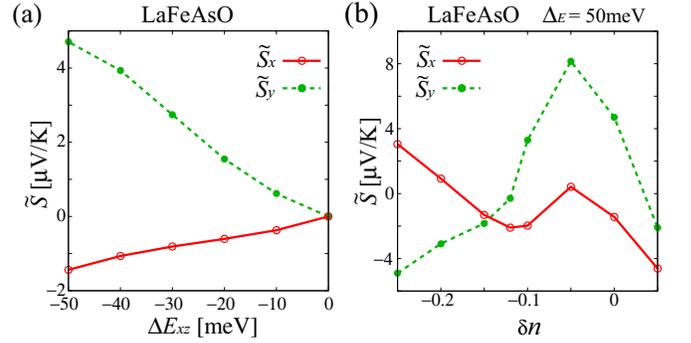}
\caption{
(a) $\Delta E_{xz}$ dependence of $\tilde{S}_{\mu}\equiv S_\mu-S^0$ in the LaFeAsO model, and
(b) carrier doping $\delta n$ dependence of $\tilde{S}_{\mu}$ for
 $\alpha_s=0.990$ and $\Delta_E=50$meV in the LaFeAsO model.
}
\label{fig4}
\end{figure}

Here, we briefly analyze the TEP $S$,
which is given as $S_{\mu}=\frac{1}{\sigma_{\mu}}
\sum_{b}\alpha^b_{\mu}$, where
{\small \begin{eqnarray}
\alpha^b_{\mu}=-\frac{e}{TN}\sum_{\k}\int_{-\infty}^{\infty}
 \frac{d\omega}{\pi}\left(-\frac{\partial f(\omega)}{\partial \omega}\right)
 \omega\left|v^{\mu}_{b,\k} G^b_{\k}(\w+i0)\right|^2 \label{alpha^b}
\end{eqnarray}
}
is the Peltier conductivity on band $b$.
Figure \ref{fig4}(a) shows the $C_2$ anisotropy of the TEP induced by the
orbital polarization in the LaFeAsO model. Here, $\tilde{S}_\mu$ is
defined as $\tilde{S}_\mu\equiv
S_\mu-S^0$, where $S^0$ is the TEP at $\Delta_E=0$meV.
The value of $\tilde{S}_y$ remarkably increases with the orbital polarization, which
is consistent with the experimental results in Eu122 \cite{Jiang}.
This result is mainly caused by the strong energy
dependence of $\gamma^b_{\k}$ near the cold spots on the h-FS2: See the SM, Sec. B \cite{SM} for
 details. We note that $S^0$ is sensitive to details of the model, 
because of the large cancellation between positive $\alpha^b$ from the
h-FSs and negative $\alpha^b$ from the e-FSs. In fact, $S^0\sim-10\mu$V/K
in the present {\it d-p} model, whereas
$S^0\sim 0$meV in the five $d$-orbital LaFeAsO model analyzed in
Ref.  \cite{Onari-Hdoped}. Nonetheless, the relations $\tilde{S}_y>0$ and
$\tilde{S}_x<0$ in Fig. \ref{fig4}(a) are robust and model-independent.
In Fig. \ref{fig4}(b),  we show the $\delta n$ dependence of $\tilde{S}_\mu$ in the
LaFeAsO model for $\Delta_E=50$meV by adjusting $r$  to satisfy $\alpha_s=0.990$. The anisotropy of
$S$ is reversed in heavily hole-doped case ($\delta n< -0.15$).

In the SM, Sec. D \cite{SM}, we also study the LaFeAsO model with the
orbital polarization only on the e-FSs, which is suggested by the ARPES measurement in Ba122 \cite{ARPES-Shen}.
The obtained anisotropies of $\rho$ and
$S$ are qualitatively the same as the case of all
FSs are polarized, since the structures of $C_2$ spin fluctuations and
$\gamma^b_{\k}$ are essentially unchanged. 
In the SM, Sec. E \cite{SM}, we study the effect of the spin-orbit
interaction (SOI) \cite{Saito-SOI} on the transport properties in FeSe.

Finally, we stress that the TEP is magnified by the mass-enhancement factor
$z^{-1}$ as shown in
Eq. (\ref{eq7}) in the SM, Sec. B \cite{SM}. The value of $z^{-1}$  observed by experiments  is
$z^{-1}\sim 3$-$5$ in EuFe$_2$As$_2$ \cite{Eu122-SdH} and
BaFe$_2$As$_2$ \cite{Ba122-dHvA,Ba122-Hashimoto,Ba122-C}. Using the experimental
$z^{-1}$, we can understand ${S}_y-{S}_x\sim 20\mu$V/K observed in
Eu122 near $T_{\rm N}$ \cite{Jiang}.


In summary, 
we studied the anisotropy in the transport coefficients 
in the nematic states
to clarify the true nematic order parameter in Fe-based superconductors \cite{FeSe-Yamakawa2,Chubukov-PRX}.
Once the orbital order sets in,
the inelastic scattering rate $\gamma^b_{\k}$ becomes very anisotropic
due to the prominent orbital-dependent spin fluctuations.
For this reason, the characteristic material-dependent $C_2$ transport
phenomena below $T_{\rm S}$ are naturally explained based on the
realistic multiorbital Hubbard models.
In particular, the opposite anisotropy $\rho_x>\rho_y$ in FeSe
originates from the singleness of the hole pocket.
In addition, the thermoelectric power shows sizable in-plane anisotropy
due to the strong energy-dependence of $\gamma^b_{\k}$.
This study leads to the conclusion that the orbital order scenario,
which is microscopically supported by the SC-VC theory,
is universal in various Fe-based superconductors.

\acknowledgements
We are grateful to Y. Yamakawa and T. Fujii for valuable discussions. 
This work was supported by JSPS KAKENHI Grant Number JP26800185.
Part of numerical calculations was
performed on the Yukawa Institute Computer Facility.



\clearpage

\makeatletter
\renewcommand{\thefigure}{S\arabic{figure}}
\renewcommand{\theequation}{S\arabic{equation}}
\makeatother
\setcounter{figure}{0}
\setcounter{equation}{0}
\setcounter{page}{1}
\setcounter{section}{1}

\begin{widetext}
\begin{center}
{\bf 
[Supplementary Material] \\
 In-plane anisotropy of transport coefficients in the electronic nematic
 states: \\
Universal origin of the nematicity in Fe-based superconductors}%
\end{center}

\begin{center}
Seiichiro Onari$^{1,2}$ and Hiroshi Kontani$^3$
\end{center}

\begin{center}
\textit{$^1$ Department of Physics, Okayama University, Okayama 700-8530, Japan}

\textit{$^2$ Research Institute for Interdisciplinary Science, Okayama
 University, Okayama 700-8530, Japan}

\textit{$^3$ Department of Physics, Nagoya University, Nagoya 464-8602, Japan}
\end{center}

\end{widetext}

\subsection{A: Details of the eight-orbital models and formulation}

\begin{figure}[!htb]
\includegraphics[width=.99\linewidth]{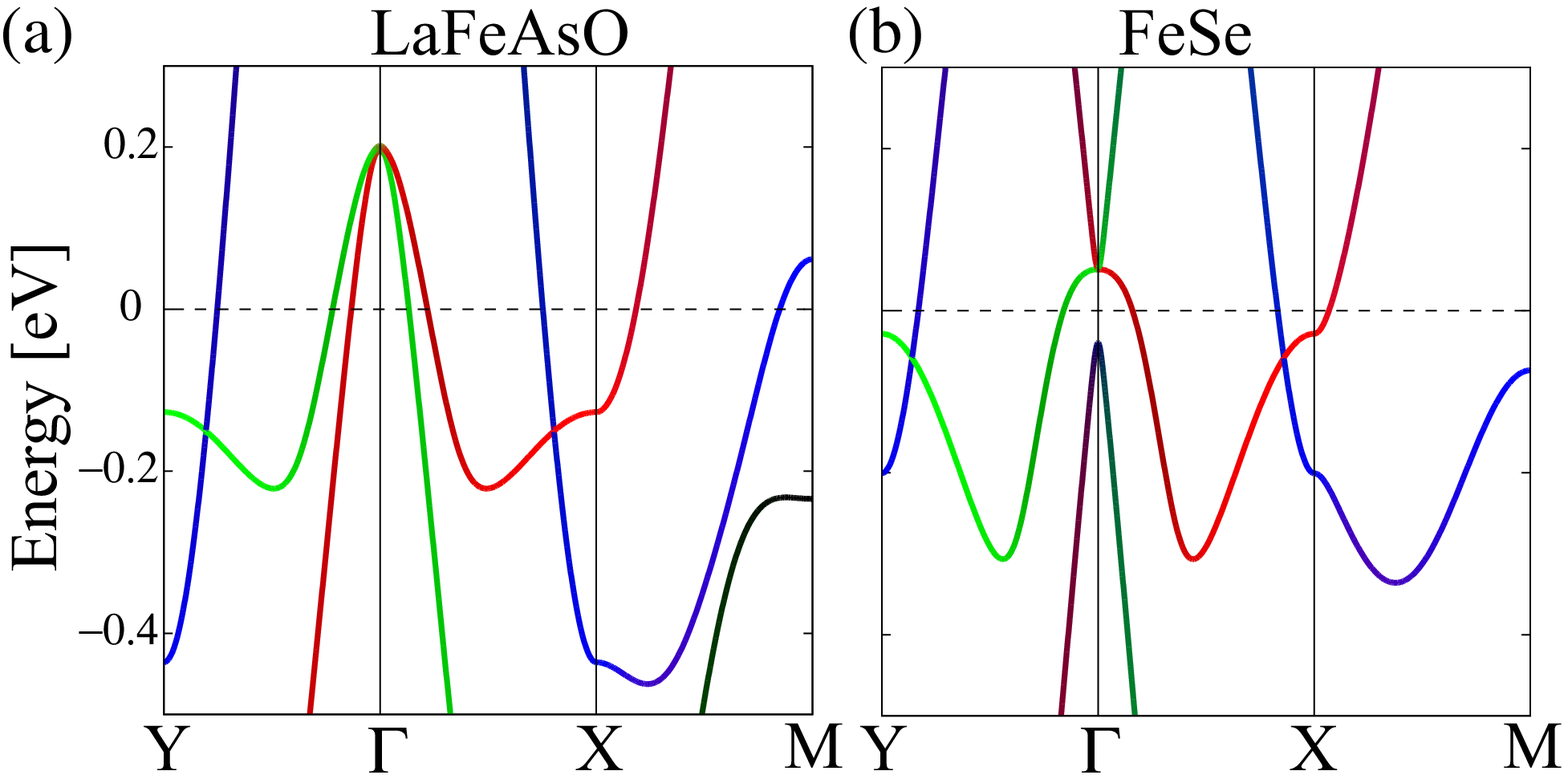}
\caption{
Band dispersions for (a) the LaFeAsO model and (b) the FeSe model. The colors correspond to 2 (green), 3 (red), and 4 (blue), respectively.
}
\label{fig:band}
\end{figure}

Here, we introduce the eight-orbital {\it d-p} models $H^0_{\rm M}$
(M$=$FeSe, LaFeAsO)
analyzed in the main text.
We first derived the first-principles tight-binding models
using the WIEN2k and WANNIER90 codes.
For FeSe, in order to obtain the experimentally observed Fermi surfaces (FSs),
we introduce the $\bm{k}$-dependent shifts for orbital $l$, $\delta E_l(\k)$,
by introducing the intra-orbital hopping parameters
as explained in Ref.  \cite{FeSe-Yamakawa2}.
We shift the $d_{xy}$-orbital band [$d_{xz/yz}$-orbital band] 
at ($\Gamma$, M, X) points
by ($-0.60$, $-0.25$, $+0.24$) [($-0.24$, $0$, $+0.12$)], in unit eV. In Figs. \ref{fig:band}(a) and \ref{fig:band}(b), we show the obtained band dispersions
for the LaFeAsO model and the FeSe model, respectively.

\begin{figure}[!htb]
\includegraphics[width=.99\linewidth]{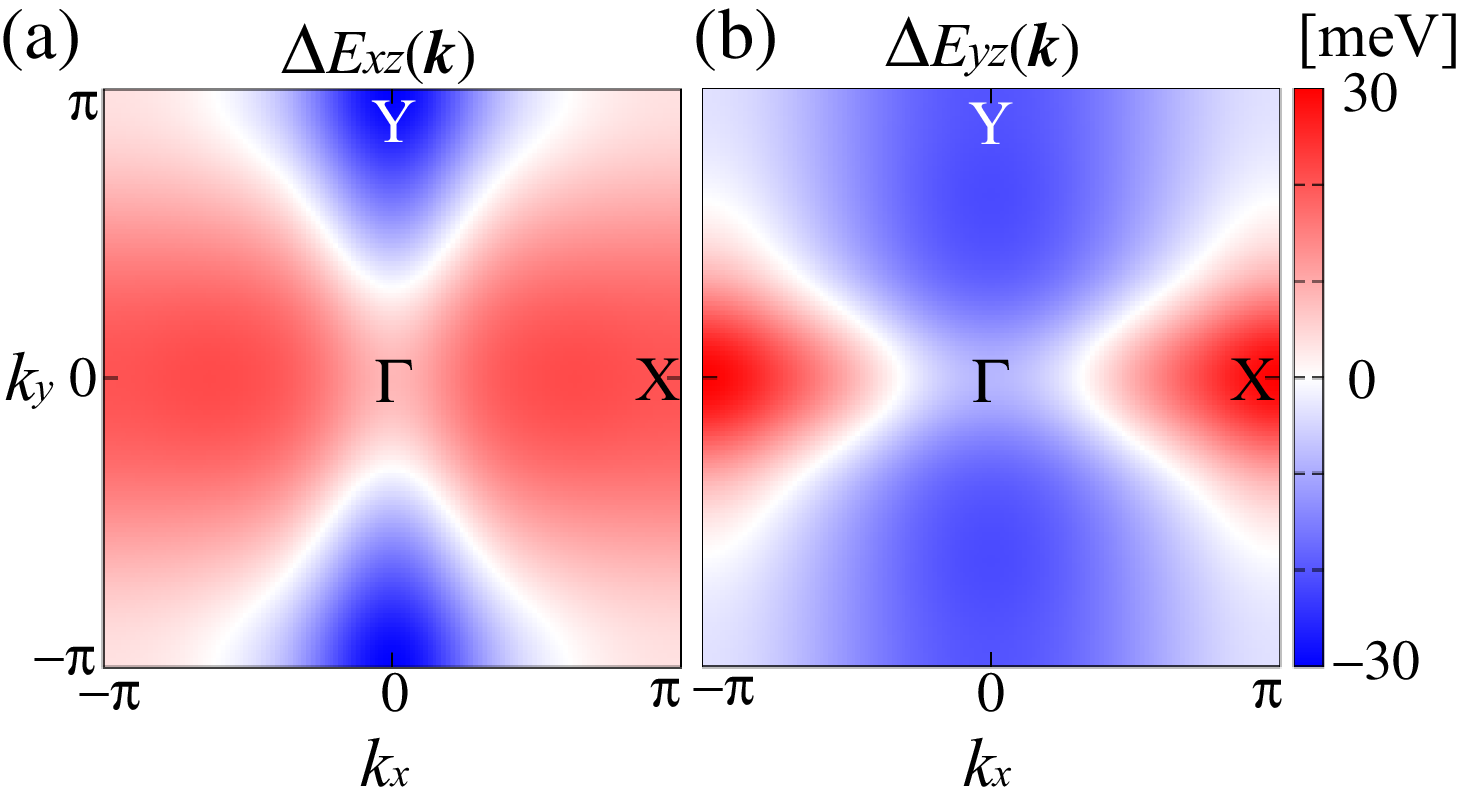}
\caption{
$\bm{k}$-dependences of the orbital polarization (a) $\Delta
E_{xz}(\bm{k})$ and (b) $\Delta
E_{yz}(\bm{k})$ for $\Delta_E=30$meV obtained by the symmetry-breaking
 self-energy method in the FeSe model \cite{Onari-FeSe}.
}
\label{fig:S0}
\end{figure}

Next, we explain the orbital polarization term $H^{\rm orb}_{\rm
M}$. For the FeSe model used in the main text is given by the symmetry-breaking self-energy method developed in
previous paper \cite{Onari-FeSe}. The obtained sign-reversing orbital polarization is shown in Fig. \ref{fig:S0}. In this orbital polarization, the
relation $\Delta E_{xz}(\Gamma)-\Delta
E_{yz}(\Gamma)>0$ and $\Delta E_{xz}({\rm Y})-\Delta
E_{yz}({\rm X})<0$ holds, consistently with the ARPES measurements \cite{FeSe-ARPES62}.

Finally, we explain the multiorbital Coulomb interaction $H^U_{\rm M}$. The bare Coulomb
interaction for the spin channel is given as
\begin{equation}
(\Gamma^{\mathrm{s}})_{l_{1}l_{2},l_{3}l_{4}} = \begin{cases}
U_{l_1,l_1}, & l_1=l_2=l_3=l_4 \\
U_{l_1,l_2}' , & l_1=l_3 \neq l_2=l_4 \\
J_{l_1,l_3}, & l_1=l_2 \neq l_3=l_4 \\
J_{l_1,l_2}, & l_1=l_4 \neq l_2=l_3 \\
0 , & \mathrm{otherwise}.
\end{cases}
\end{equation}
Also, the bare Coulomb interaction for the charge channel is
\begin{equation}
({\hat \Gamma}^{\mathrm{c}})_{l_{1}l_{2},l_{3}l_{4}} = \begin{cases}
-U_{l_1,l_1}, & l_1=l_2=l_3=l_4 \\
U_{l_1,l_2}'-2J_{1_1,l_2} , & l_1=l_3 \neq l_2=l_4 \\
-2U_{l_1,l_3}' + J_{l_1,l_3} , & l_1=l_2 \neq l_3=l_4 \\
-J_{1_1,l_2} , &l_1=l_4 \neq l_2=l_3 \\
0 . & \mathrm{otherwise}.
\end{cases}
\end{equation}
Here, $U_{l,l}$, $U_{l,l'}'$ and $J_{l,l'}$
are the first-principles Coulomb interaction terms
given in Ref.  \cite{Arita2}.
The interaction matrix
for the self-energy $\hat{V}^\Sigma$ is given as \cite{Onari-Hdoped,Onari-FeSe,Onari-Springer}
\begin{eqnarray}
\hat{V}^\Sigma(q)&=&\frac{3}{2}\hat{\Gamma}^s\hat{\chi}^s(q)\hat{\Gamma}^s+\frac{1}{2}\hat{\Gamma}^c\hat{\chi}^c(q)\hat{\Gamma}^c
 \nonumber \\
&&-\frac{1}{4}(\hat{\Gamma}^c-\hat{\Gamma}^s)\hat{\chi}^{\rm
 irr}(q)(\hat{\Gamma}^c-\hat{\Gamma}^s)  \nonumber \\
&&-\frac{1}{8}(\hat{\Gamma}^c+\hat{\Gamma}^s)\hat{\chi}^{\rm
 irr}(q)(\hat{\Gamma}^c+\hat{\Gamma}^s).
\end{eqnarray}

\subsection{B: Origin of the large in-plane anisotropy of $S$ in the
  LaFeAsO model}
In the following, we explain the reason why the in-plane anisotropy of $S$
becomes large with increasing $\Delta_E$ in the LaFeAsO model.
$\alpha^b_{\mu}$ introduced in Eq. (\ref{alpha^b}) in the main text is
rewritten as
{\small\begin{eqnarray}
\!\!\!\!\alpha^b_{\mu}\!\!&=&\!\!-\frac{e}{T}\int_{\rm FS} \frac{dk^b_\parallel}{(2\pi)^2}
 \int \frac{d\epsilon^{b*}_{\bm{k}}}{|v_{b,\k}|}\left(-\frac{\partial f}{\partial \epsilon}\right)_{\epsilon=\epsilon^{b*}_{\bm{k}}}\!\!\!\!
\frac{\epsilon^{b*}_{\bm{k}}|v^{\mu}_{b,\k}|^2}{\gamma^b_{\bm{k}}}
\label{eq7-1}\\
\!\!&\approx&\!\!-\frac{e\pi^2T}{3}\int_{\rm FS}
 \frac{dk^b_\parallel}{(2\pi)^2}\frac{1}{z^b_{\bm{k}}|v_{b,\k}|}\frac{\partial}{\partial
 k^b_\perp}\!\!\left(\frac{|v^{\mu}_{b,\k}|^2}{|v_{b,\k}|\gamma^b_{\bm{k}}}\right)\!\!,\label{eq7}
\end{eqnarray}
}where $k^b_\parallel$ and $k^b_\perp$ denote $\bm{k}$ along the FS and
$\bm{k}$ perpendicular to the FS on band $b$, respectively. $\epsilon^{b*}_{\bm{k}}$ is the
renormalized quasiparticle energy given by
$\epsilon^{b*}_{\bm{k}}=z^{b}_{\k}[\varepsilon^{b}_{\bm{k}}+{\rm Re}\Sigma^b(\bm{k},0+i0)-\mu]$, and
$\gamma^b_{\bm{k}}=-{\rm
Im}\Sigma^{b}(\bm{k},\epsilon^{b*}_{\bm{k}}+i0)$ is the quasiparticle
damping without renormalization. The
mass renormalization factor $z^{b}_{\k}$ is given by
$z^{b}_{\k}=\left[1-\frac{\partial {\rm Re}\Sigma^{b}(\bm{k},\w+i0)}{\partial
\w}|_{\w=0}\right]^{-1}$. 
According to Eq. (\ref{eq7-1}), $\a^b_{\mu}$ is sensitively influenced
by the $\epsilon^{b*}_{\bm{k}}$ dependence of
$1/\gamma^b_{\bm{k}}$, and $1/\gamma^b_{\bm{k}}$ is strongly
energy-dependent in correlated electron systems. For instance, $\a^b_{\mu}\sim0$ is obtained when
$1/\gamma^b_{\bm{k}}$ is symmetric with respect to $\epsilon^{b*}_{\bm{k}}\rightarrow-
\epsilon^{b*}_{\bm{k}}$ since
$\epsilon^{b*}_{\bm{k}}\left(-\frac{\partial f}{\partial
\epsilon}\right)_{\epsilon=\epsilon^{b*}_{\bm{k}}}$ is an odd function of $\epsilon^{b*}_{\bm{k}}$.

\begin{figure}[!htb]
\includegraphics[width=\linewidth]{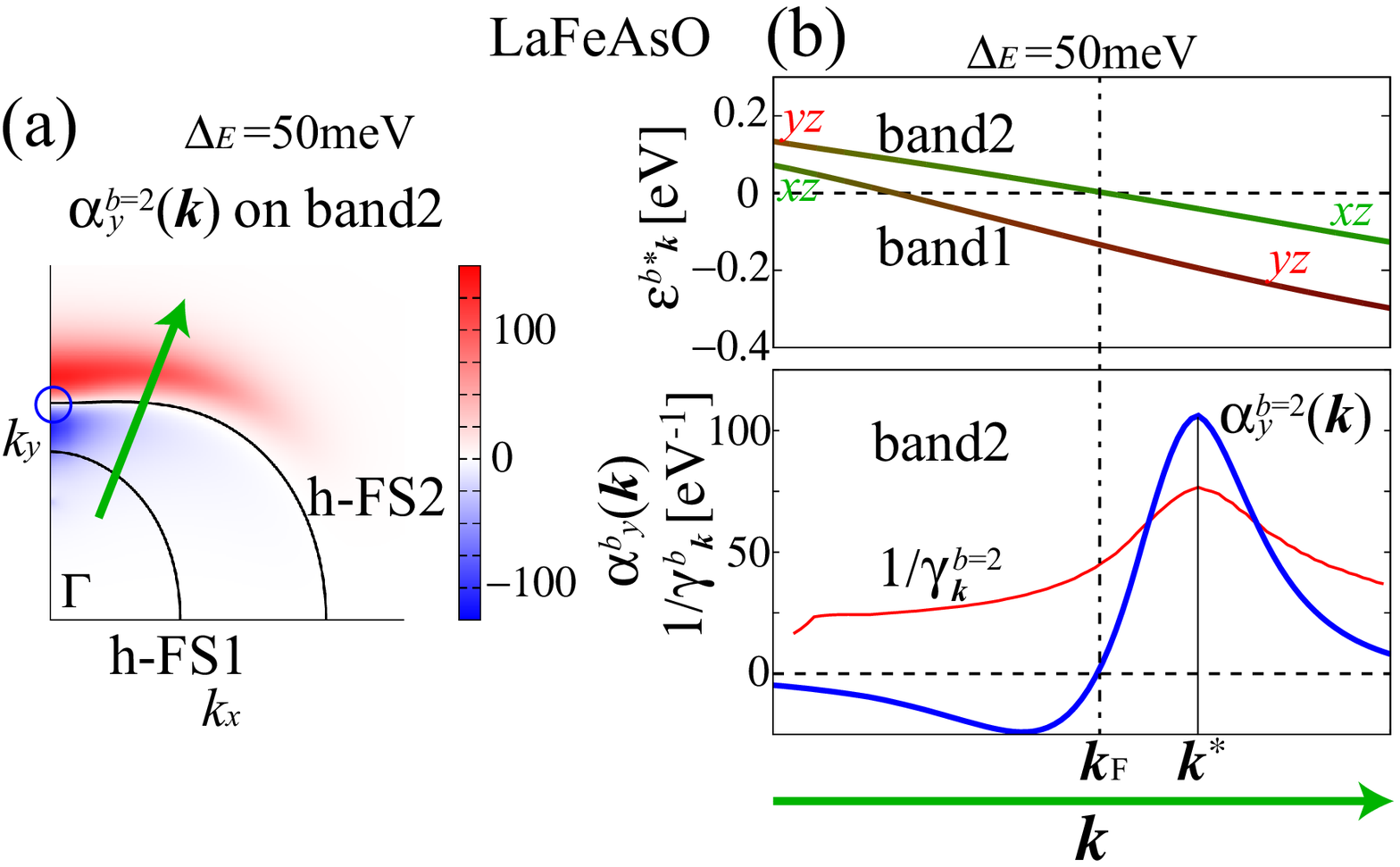}
\caption{
(a) $\bm{k}$ dependence of $\alpha^{b=2}_{y}(\bm{k})$ around h-FS2
in the LaFeAsO model for $\Delta_E=50$meV. (b) $\bm{k}$ dependences of
 $\epsilon^{b*}_{\bm{k}}$ for $b=1$ and $2$ in the upper panel, $\alpha^{b=2}_{y}(\bm{k})$ and $1/\gamma^{b=2}_{\bm{k}}$ in the lower panel, as functions of $\k$ along the green arrow in (a). 
}
\label{fig:S-reason}
\end{figure}

Here, we introduce $\a^b_{\mu}(\bm{k})$ as 
{\small \begin{eqnarray}
\alpha^b_{\mu}(\bm{k})=-\frac{e}{T}\int_{-\infty}^{\infty}
 \frac{d\omega}{\pi}\left(-\frac{\partial f(\omega)}{\partial \omega}\right)
 \omega\left|v^{\mu}_{b,\k} G^b_{\k}(\w+i0)\right|^2.
\end{eqnarray}
}
Then, the Peltier conductivity for band $b$ is $\a^b_{\mu}=\frac{1}{N}\sum_{\k}\a^b_{\mu}(\bm{k})$.
In Fig. \ref{fig:S-reason}(a), we show the obtained $\bm{k}$ dependence of
$\a^{b=2}_{y}(\bm{k})$ on band2 including the h-FS2 around the $\Gamma$
point in the LaFeAsO model for $\Delta_E=50$meV. $\a^{b=2}_{y}(\bm{k})$ has large value around the cold
spots, and the area for positive $\a^{b=2}_{y}(\bm{k})$ is much wider
than the area for negative $\a^{b=2}_{y}(\bm{k})$. This result originates from 
the highly asymmetric $\bm{k}$ dependence of
$1/\gamma^{b=2}_{\bm{k}}$ near the Fermi momentum.
In Fig. \ref{fig:S-reason}(b), we show $\epsilon^{b*}_{\bm{k}}$ for
$b=1$ (h-FS1) and $b=2$ (h-FS2) in the upper panel, and 
$1/\gamma^{b=2}_{\bm{k}}$ and $\a^{b=2}_{y}(\bm{k})$ on the band2 in the lower
panel, as functions of $\k$ along the green arrow illustrated in
Fig. \ref{fig:S-reason}(a). 
We see that the positive value of $\a^{b=2}_{y}(\bm{k})$ is much larger than
the negative value of $\a^{b=2}_{y}(\bm{k})$ in magnitude.
In addition, both $\a^{b=2}_{y}(\bm{k})$ and $1/\gamma^{b=2}_{\bm{k}}$
take the maxima at $\k=\k^*$.
Thus, the large positive $\tilde{S}_y$ originates from the strong asymmetry
of $1/\gamma^b_{\bm{k}}$ near the Fermi surface \cite{Kontani-S2}.
The asymmetry of
$1/\gamma^b_{\bm{k}}$ is caused by the orbital
dependence of $\gamma^b_{\k}$. In the orbital basis, we explain in the
main text that the
quasiparticle damping for the
$d_{yz}$ orbital is much larger than that for the $d_{xz}$
orbital $(\gamma_{yz}\gg \gamma_{xz})$ since the spin fluctuations
develop mainly on the $d_{yz}$ orbital. As shown by the colors on
the band dispersion in Fig. \ref{fig:S-reason}(b), 
$d_{xz}$ orbital is dominant for $\k\approx\k^*$, and weight of $d_{yz}$
orbital increases as $\k$ approaches to 
the $\Gamma$ point. Thus, the asymmetric energy dependence of
$1/\gamma^b_{\bm{k}}$ stems from the suppression 
by $\gamma_{yz}$.
On the other hand, $S_x$ slightly decreases with increasing $\Delta_E$ mainly due
to the contribution from the cold spots on the e-FSs.

\subsection{C: Carrier doping dependence of the in-plane
  anisotropies in $\rho$ and $S$ in the LaFeAsO model}
\begin{figure}[!htb]
\includegraphics[width=.99\linewidth]{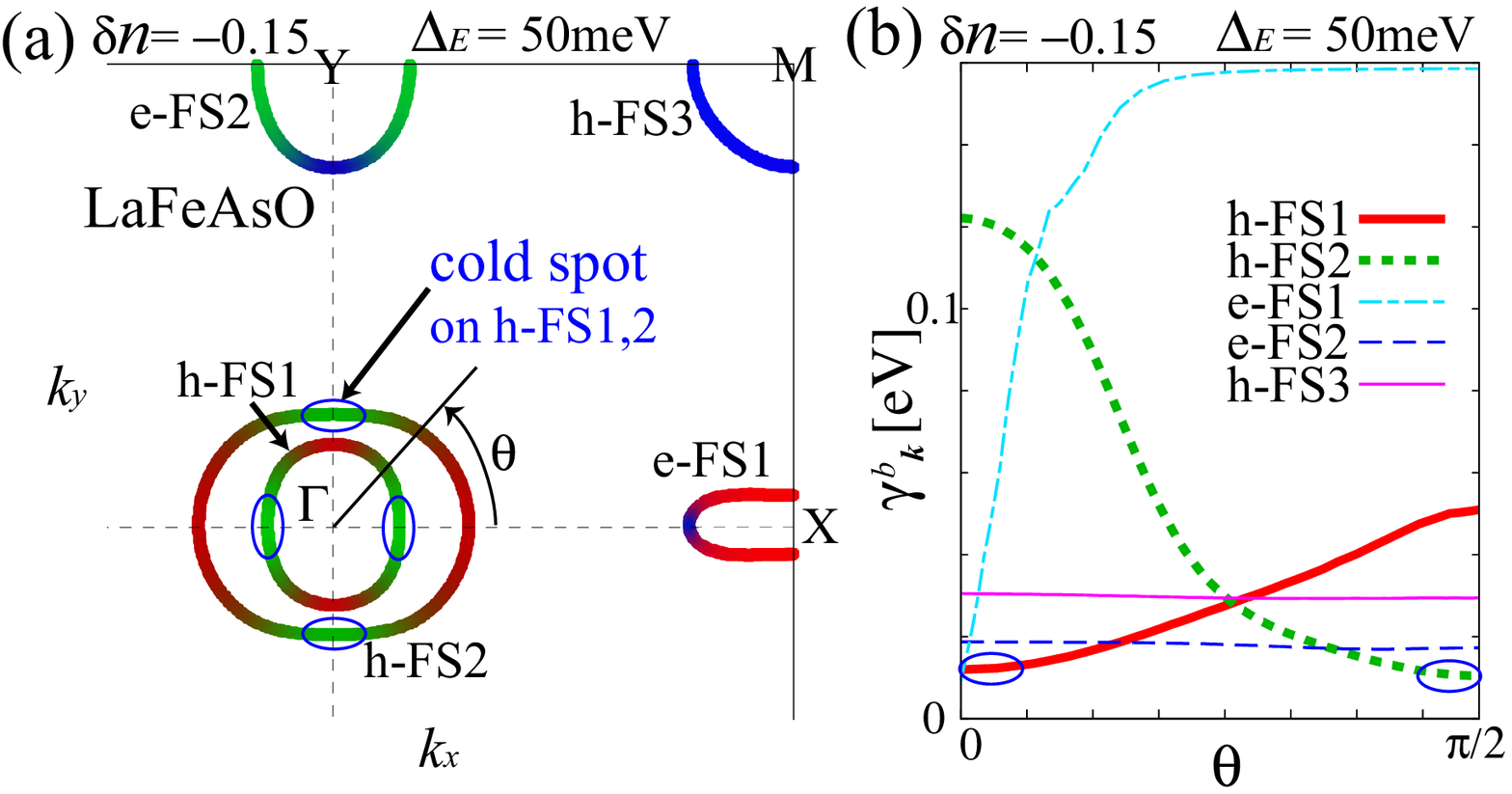}
\caption{
(a) FSs for heavily hole-doped LaFeAsO model ($\delta n=-0.15$) for
 $\Delta_E=50$meV. (b) Obtained $\theta$ dependences of
 $\gamma^b_{\k}$
}
\label{fig:n-dep}
\end{figure}

Here, we study the carrier doping $\delta n$ dependence of the
in-plane anisotropies in $\rho$ and $S$. In the hole-doped compounds
Ba$_{1-x}$K$_x$Fe$_2$As$_2$, $\rho_x$ is slightly larger than $\rho_y$
 \cite{hole-doped122,hole-doped122-2},
which is opposite to the relation $\Delta\rho<0$
observed in
the non-doped and the electron-doped Ba122.
In Fig. \ref{fig3}(f) in the main text, we show the $\delta n$
dependence of $\rho_{\mu}$ for $\Delta_E=50$meV in the LaFeAsO
model. $\a_s$ is set as $0.990$. The
 obtained sign reversal in the
 hole-doped region $(\delta n <-0.12)$ is consistent with
experimental results in the hole-doped Ba$_{1-x}$K$_x$Fe$_2$As$_2$ \cite{hole-doped122,hole-doped122-2}.
In the hole-doped LaFeAsO model, the FSs and the cold spots are shown in
Fig. \ref{fig:n-dep}(a). The relation $\Delta\rho>0$ is mainly
originates from the h-FS2, since the anisotropy of $\gamma$ on the h-FS2 is
larger than that on the h-FS1 as shown in Fig. \ref{fig:n-dep}(b).

In Fig. \ref{fig4}(b) in the main text, we also show the $\delta n$ dependences of $\tilde{S}_{\mu}$
 for $\Delta_E=50$meV in the LaFeAsO model. $\a_s$ is set as $0.990$. We
 obtain the
 reverse of the in-plane anisotropy ($S_x>S_y$) in heavily
 hole-doped case $(\delta n<-0.15)$. This reversal is caused by the
 competition between the contribution from the h-FS1 and that from the h-FS2:
The h-FS1 contributes to the relation $S_x>S_y$, while the h-FS2
contributes to the opposite relation $S_x<S_y$.
The former contribution becomes larger than the latter contribution in
 hole-doped case ($\delta n <-0.15$).
We note that the contribution from the e-FSs is unimportant for the anisotropies of
$\rho$ and $S$, since the area of cold spot on the
e-FS1 is very narrow and $\gamma^b_{\k}$ on the e-FS2 is almost
 isotropic as shown in Fig. \ref{fig:n-dep}(b).


\subsection{D: Orbital polarization only on the electron FSs in the LaFeAsO model}
\begin{figure}[!htb]
\includegraphics[width=.99\linewidth]{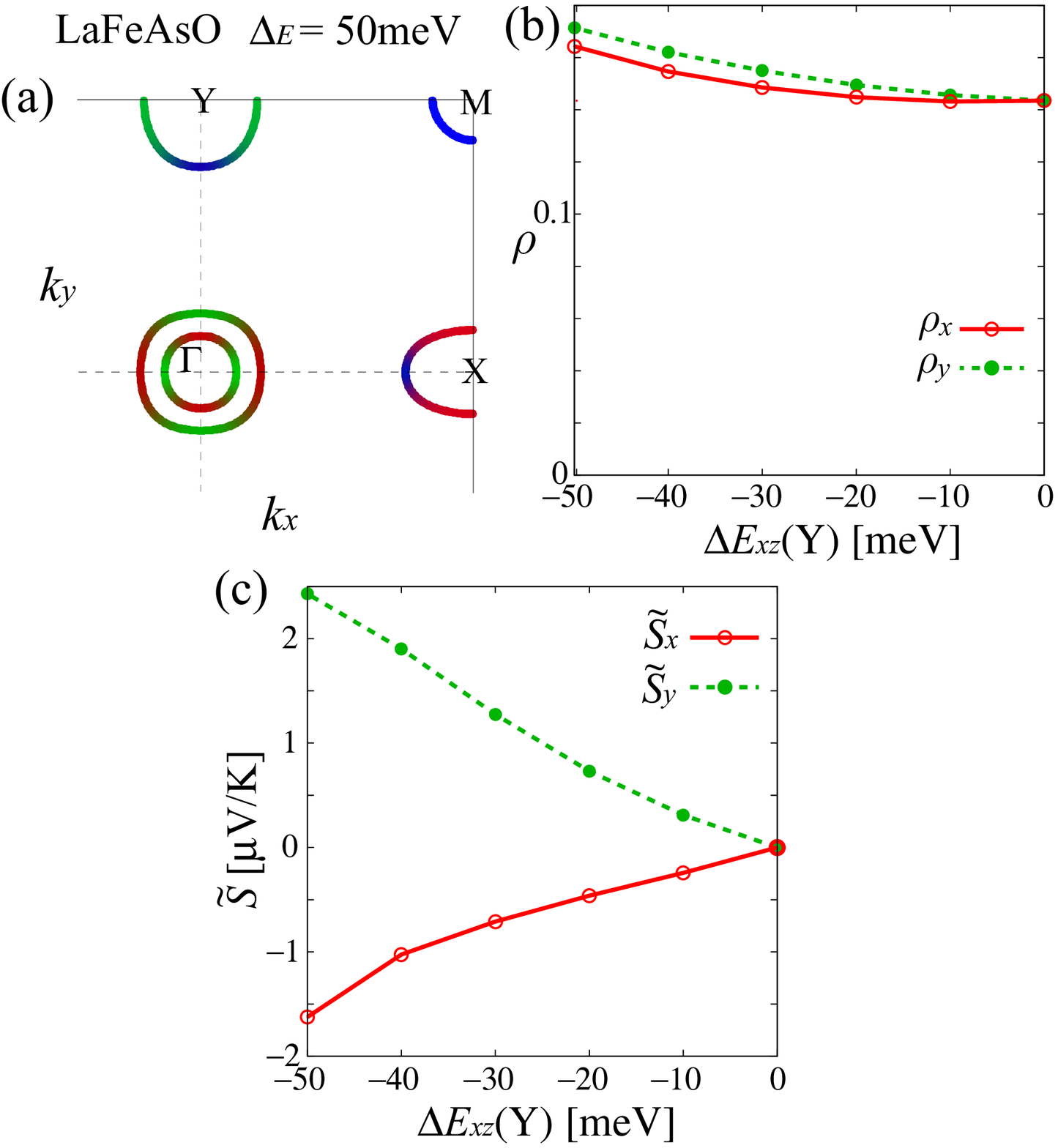}
\caption{
(a) FSs of the LaFeAsO model for
 $\Delta_E=50$meV only on the e-FSs ($\alpha_s=0.983$).
$\Delta E_{xz}$(Y) dependences of (b) $\rho_{\mu}$ and (c) $\tilde{S}_{\mu}$.
}
\label{fig:S1}
\end{figure}

In the main text, we employed the constant orbital polarization $\Delta
E_{xz}(\bm{k})=-\Delta_E$, $\Delta
E_{yz}(\bm{k})=\Delta_E$ in the LaFeAsO model.
In order to verify the validity of the results obtained in the main
text, here we introduce the orbital polarization $\Delta
E_{xz}(\bm{k})=-\Delta E_{yz}(\bm{k})=-\Delta_E$ only around the X, Y
points whereas $\Delta
E_{xz}(\Gamma)=\Delta E_{yz}(\Gamma)=0$. Such $\k$-dependent orbital
polarization has been reported by the ARPES measurement in BaFe$_2$As$_2$ \cite{ARPES-Shen}.
In Fig. \ref{fig:S1}(a), we show the FSs for $\Delta_E=50$meV. For
$r=0.334$, the obtained $\rho_{\mu}$ and
$\tilde{S}_{\mu}$ as functions of  $\Delta E_{xz}$(Y) are shown in
Figs. \ref{fig:S1}(b) and \ref{fig:S1}(c), respectively.
The obtained anisotropies of $\rho$ and $S$ are essentially similar to those in Figs. \ref{fig3}(a) and \ref{fig4}(a) in the main text.

%
%
%

\subsection{E: Results including the effect of the SOI in FeSe}
In the main text, the spin-orbit interaction (SOI) is not taken into account.
Here, we study the effect of the SOI,
which is expressed as 
$\lambda \sum_i {\bm l}_i\cdot {\bm \sigma}_i$.
The matrix elements of ${\bm l}_i$ are given in Ref.  \cite{Saito-SOI}
.
In the presence of the SOI, we have to study the sixteen-orbital model
in the folded Brillouin zone (BZ) picture
since the ``unfolding'' is prohibited by the SOI.
Since the numerical calculation becomes heavy in the presence of the SOI,
 we take smaller $N=N_x\times N_y=64\times64$ $\bm{k}$ meshes 
and 512 Matsubara frequencies compared to the main text.

\begin{figure}[!htb]
\includegraphics[width=.99\linewidth]{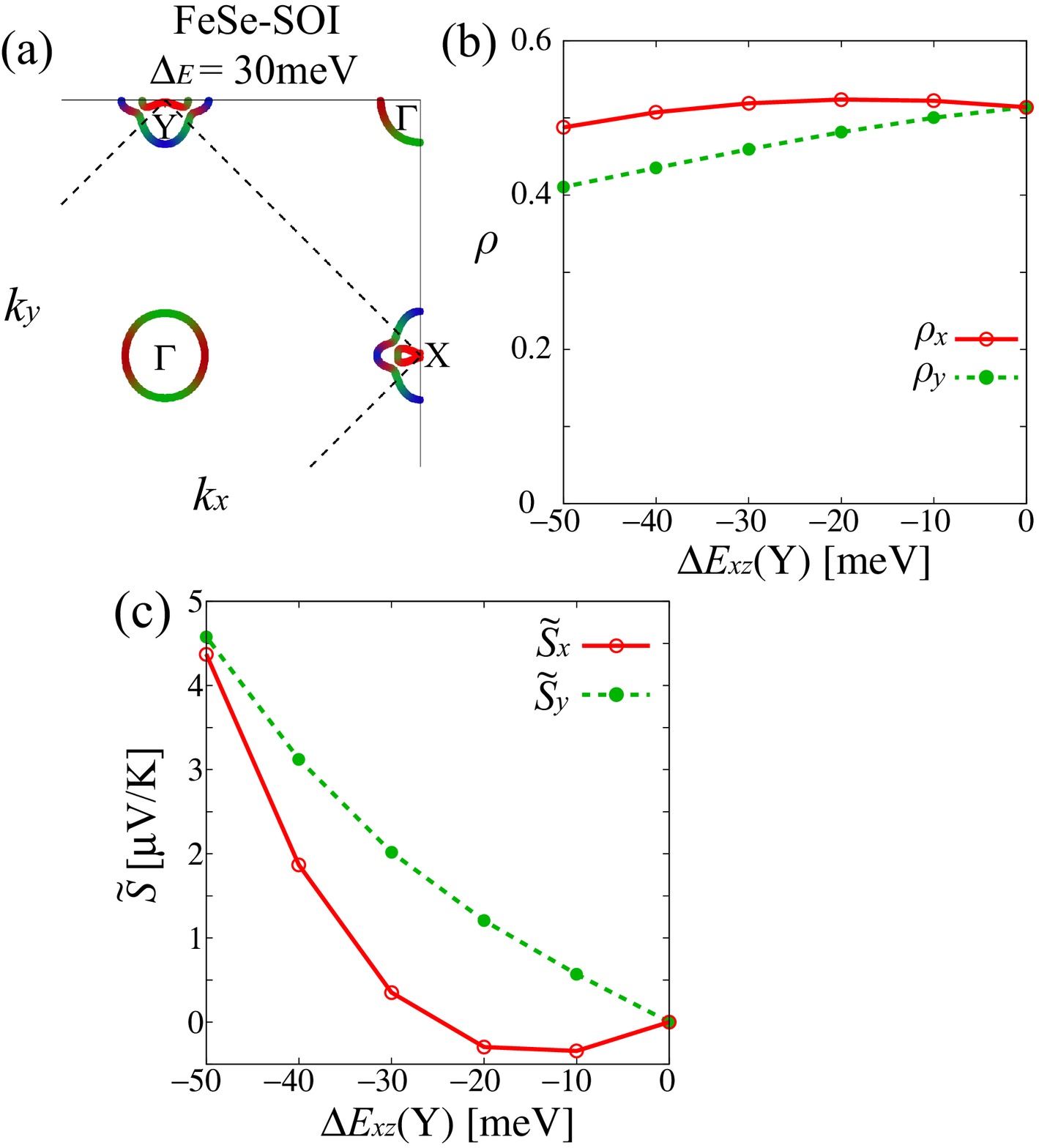}
\caption{
(a) FSs of the 16 orbital {\it d-p} FeSe model for
 $\Delta_E=30$meV and SOI $\lambda=50$meV ($\alpha_s=0.865$).
 $\Delta E_{xz}$(Y) dependence of (b) $\rho_{\mu}$ and (c) $\tilde{S}_{\mu}$ for $\lambda=50$meV.
}
\label{fig:S3}
\end{figure}

In Fig. \ref{fig:S3}(a), we show the FSs for FeSe in the folded BZ (dotted
line) for the SOI $\lambda=50$meV and $\Delta_E=30$meV. The employed
$\Delta E_{xz(yz)}(\bm{k})$ is the same as that employed in the
main text.
We put $r=0.225$. In this case $\alpha_s=0.870$ is satisfied for $\Delta_E=50$meV.
The obtained $\rho_{\mu}$ 
is shown in
Figs. \ref{fig:S3}(b) as a function of $\Delta_{xy}$(Y).
The obtained result is qualitatively the same as the results without
the SOI shown in Fig. \ref{fig3} in the main text.
In Fig. \ref{fig:S3}(c), we show the obtained $\tilde{S}_{\mu}$ as a function of $\Delta_{xy}$(Y).
The obtained anisotropy of $S$ is small because of
the nearly symmetric energy dependence of
$1/\gamma^{b=2}_{\bm{k}}$ due to the moderate spin fluctuations in FeSe.

\begin{thebibliography}{99}
\bibitem{FeSe-model}
In the present FeSe model, we shift $E_4(\k)$ around the $\Gamma$ point
	by $-0.6$eV in the original FeSe model in Ref.  \cite{FeSe-Yamakawa2} in order to realize the single hole-pocket.
\end{thebibliography}
\end{document}